\documentclass{article}
\usepackage{ismir,amsmath,cite,url}
\usepackage[usenames,dvipsnames,svgnames,table]{xcolor}

\usepackage{textcomp} 
\usepackage{booktabs}

\usepackage{subfiles}

\usepackage{rotating}
\usepackage{graphicx}
\usepackage{caption}
\usepackage{subcaption}
\usepackage{makecell}
\usepackage{soul} 
\usepackage{diagbox} 

\usepackage{mathtools}
\usepackage{float}
\usepackage{amsmath}

\DeclareMathOperator*{\argmax}{arg\,max}

\renewenvironment{description}[1][0pt]
  {\list{}{\labelwidth=.25cm \leftmargin=#1
   }}
  {\endlist}

\graphicspath{{./figs/}}

\newcommand{\WASABI}[0]{WASABI }
\newcommand{\MASK}[1]{}

\title{DALI: a large Dataset of synchronized Audio, LyrIcs and notes, automatically created using teacher-student machine learning paradigm}

\multauthor
  {Gabriel Meseguer-Brocal \hspace{1cm} Alice Cohen-Hadria \hspace{1cm} Geoffroy Peeters}
  {
    \small Ircam Lab, CNRS, Sorbonne Universit\'{e}, Minist\`{e}re de la Culture, F-75004 Paris, France \\
    {\tt\small  gabriel.meseguerbrocal@ircam.fr, alice.cohenhadria@ircam.fr, geoffroy.peeters@ircam.fr}
  }
\def\authorname{Gabriel Meseguer-Brocal, Alice Cohen-Hadria, Geoffroy Peeters}

\begin{document}

  \maketitle

  \begin{abstract}

    The goal of this paper is twofold.
    First, we introduce DALI, a large and rich multimodal dataset containing 5358 audio tracks with their time-aligned vocal melody notes and lyrics at four levels of granularity.

    The second goal is to explain our methodology where dataset creation and learning models interact using a teacher-student machine learning paradigm that benefits each other.
    We start with a set of manual annotations of draft time-aligned lyrics and notes made by non-expert users of Karaoke games.
    This set comes without audio.
    Therefore, we need to find the corresponding audio and adapt the annotations to it.
    To that end, we retrieve audio candidates from the Web.
    Each candidate is then turned into a singing-voice probability over time using a teacher, a deep convolutional neural network singing-voice detection system (SVD), trained on cleaned data.
    Comparing the time-aligned lyrics and the singing-voice probability, we detect matches and update the time-alignment lyrics accordingly.
    From this, we obtain new audio sets. They are then used to train new SVD students used to perform again the above comparison.
    The process could be repeated iteratively.
    We show that this allows to progressively improve the performances of our SVD and get better audio-matching and alignment.

  	%
  	%

  \end{abstract}

  \section{Introduction}
  Singing voice is one of the most important elements in popular music.
  It combines its two main dimensions: melody and lyrics.
  Together, they tell stories and convey emotions improving our listening experience.
  Singing voice is usually the central element around which songs are composed.
  It adds a linguistic dimension that complements the abstraction of the musical instruments.
  The relationship between lyrics and music is both \textit{global} (lyrics topics are usually highly related to music genre) and \textit{local} (it connects specific musical parts with a concrete lexical meaning, and also defines the structure of a song).

  Despite its importance, singing voice has not received much attention from the MIR community.
  It has only been introduced a few years ago as a standalone topic\cite{Goto_2014, Mesaros_2013}.
  One of the most important factors that prevents its development is the absence of large and good quality reference datasets.
  This problem also exists in other MIR fields, nevertheless several solutions have been proposed\cite{Benzi_2016, Fonseca_2017}.
  Currently, researchers working in singing voice use small designed dataset following different methodology\cite{Fujihara_2012}.
  Large datasets as the one used in \cite{Humphrey_2017} are private and not accessible to the community.

  The goal of this paper is to propose such a dataset and to describe the methodology followed to construct it.

  \begin{figure*}[ht]
  	\centerline{
  		\includegraphics[width=\textwidth]{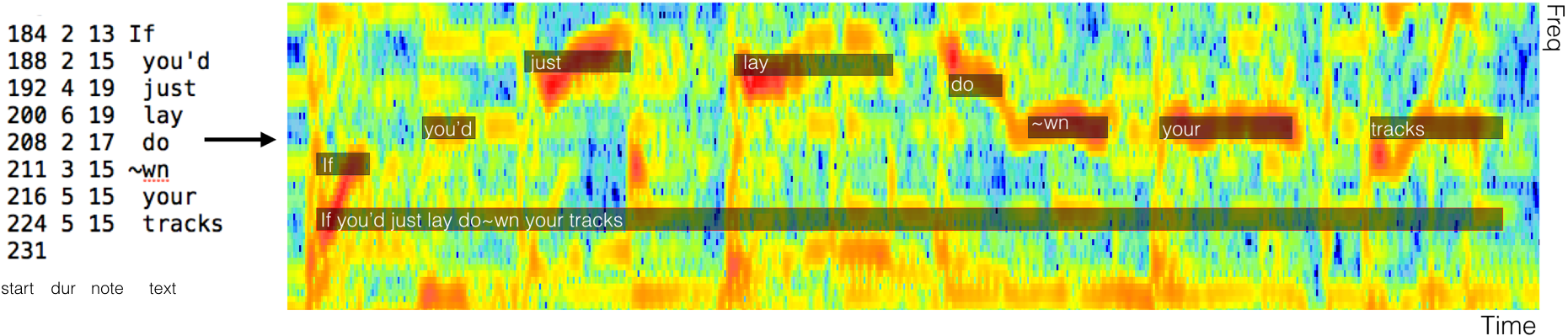}
  	}
  	\caption{[Left part] The inputs of our dataset creation system are karaoke-user annotations presented as a triple of \{time (start + duration), musical-notes, text\}.
  		[Right part] Our dataset creation system automatically finds the corresponding full-audio track and aligned the vocal melody and the lyrics to it.
  		In this example, we illustrate the alignment for a small excerpts.
  		We only represent two levels of lyrics granularity: notes and lines.}
  	\label{fig:ex_annotation}
  \end{figure*}

  \subsection{Proposal}
  %

  We present the \textbf{DALI} dataset: a large \textbf{D}ataset of synchronised \textbf{A}udio, \textbf{L}yr\textbf{I}cs and notes that aims to stand as a reference for the singing voice community.
  It contains 5358 songs (real music) each with -- its audio in full-duration, -- its time-aligned lyrics and -- its time-aligned notes (of the vocal melody).
  Lyrics are described according to four levels of granularity: notes (and textual information underlying a given note), words, lines and paragraphs.
  For each song, we also provide additional multimodal information such as genre, language, musician, album covers or links to video clips.
  The rest of this paper focuses on our methodology for creating DALI.
  In Figure~\ref{fig:ex_annotation}, we illustrate the input and output of our dataset creation system.
  See Section \ref{sec:dataset_description} for more details about the dataset itself.

  The DALI dataset has been created automatically.
  Our approach consists in a constant interaction between dataset creation and learning models where they benefit from each other.
  We developed a system that acquires lyrics and notes aligned in time and finds the corresponding audio tracks.
  The time-aligned lyrics and notes come from Karaoke resources (see Section \ref{sec:karaoke} for more deatils).
  Here, non-expert users manually describe the lyrics of a song as a sequence of annotations: time aligned notes with their associated textual information.
  While this information is powerful it has two major problems:
  \textbf{1)} there is no information about the exact audio used for the annotation process (only the song title and artist name which may lead to many different audio versions),
  \textbf{2)} even if the right audio is found, annotations may need to be adjusted to fit the audio perfectly.
  In Section \ref{sec:linking_auido}, we define how we retrieve from the Web the possible audio candidates for each song.
  In Section \ref{sec:selecting}, we describe how we select the right audio among all the possible candidates and how we automatically adapt the annotated time-alignment lyrics to this audio.
  In order to do this, we propose a distance that measures the correspondence between an audio track and a sequence of manual annotations.
  This distance is also used to perform the necessary adaptations on the annotations to be perfectly aligned with the audio.
  Our distance requires the audio to be described as a singing voice probability sequence.
  This is computed using a singing voice detection (SVD) system based on deep convolutional neural network (ConvNet).
  The performance of our system highly depends on the precision of the SVD.
  Our first version is trained on few but accurately-labeled ground truths.
  While this system is sufficient to select the right audio it is not to get the best alignment.
  To improve the SVD, in Section \ref{sec:teacher-student} we propose to use a teacher-student paradigm.
  Thanks to the first SVD system (the teacher) we selected a first set of audio tracks and their corresponding annotations.
  Using them, we train new SVD systems (the students).
  We show in Section \ref{sec:experiments} that new SVD systems (the students) are better than the initial one (the teacher).
  With this new version, we increase the quality and size of the DALI dataset.
  Finally, we discuss our research in Section \ref{sec:discussion}.


  \section{Related works}
  \label{sec:related_works}

  \begin{table*}[t]
  	\centering
  	\caption{Terms overview: definition of each term used in this paper.}
  	\label{table:terms}
  	\footnotesize
  	\begin{tabular}{ r  l }

  		Term & Definition  \\
  		\toprule
  		\scriptsize{\tt Annotation} & basic alignment unit as a triple of time (start + duration wrt $Fr$), musical-notes (with 0 = C3) and text. \\
  		\scriptsize{\tt  A file with annotations}  & group of annotations that define the alignment of a particular song. \\
  		\scriptsize{\tt Offset\_time} as $O$  & it indicates the start of the annotations, its modifications moves all bock to the right or left. \\
  		\scriptsize{\tt Frame rate} as $Fr$ & it controls the annotation grid size stretching or compressing its basic unit. \\
  		\scriptsize{\tt Annotation voice sequence}  as $avs(t) \in \{0,1\}$ & singing voice (SV) sequence extracted from karaoke-users annotations. \\
  		\scriptsize{\tt  Predictions} as $\hat{p}(t) \in [0,1]$ & SV probability sequence provided by our singing voice detection. \\
  		\scriptsize{\tt  Labels}  & labels sequence of well known ground truths datasets checked by the MIR community. \\
  		\scriptsize{\tt  Teacher}  & first SV detection (SVD) system trained on \scriptsize{\tt  Labels}. \\
  		\scriptsize{\tt Student}  & new SVD system trained on the $avs(t)$ for the subset of track for which $NCC(\hat{o}, \hat{fr}) \geq T_{corr}$. \\
  	\end{tabular}
  \end{table*}

  We review previous works related to our work: singing voice detection methods and the teacher-student paradigm.\newline

  \textbf{Singing Voice detection.}
  Most approaches share a common architecture.
  Short-time observations are used to train a classifier that discriminates observations (per frame) in vocal or non-vocal classes.
  The final stream of predictions is then post-processed to reduce artifacts.

  Early works explore classification techniques such as Support Vector Machines (SVMs) \cite{Ramona_2008, Mauch_2011}, Gaussian mixture model (GMM) \cite{Fujihara_2011} or multi-layer perceptron (MLP)\cite{Berenzweig_2002}.
  Other approaches also tried to use specific vocal traits such as vibrato and tremolo \cite{Regnier_2009} or to adapt speech recognition systems for the particularities of singing voice \cite{Berenzweig_2001}.
  Over the past few years, most works focus on the use of deep learning techniques.
  For example, \cite{Schluter_2015, Schluter_2016} propose the use of ConvNet combined with data augmentation techniques (to increase the size of the training set) or trained on weakly labeled data (the data are only labeled at the file level, not at the segment level).
  \cite{Humphrey_2017} also proposes the use of CNN but with a Constant-Q input and a training on a very large private datasets mined from Spotify resources.
  Some researchers suggest the use of Recurrent Neural Networks (RNN)\cite{Lehner_2015} or Long Short-Term Memory (LSTM) \cite{Leglaive_2015}.
  One advantage of these models is that they directly model the decisions sequence over time and no post-processing is needed.
  Other singing voice detection systems are developed to be used as a pre-processing-step: for lyrics transcription \cite{Mesaros_2013} or for  source separation \cite{Simpson_2015} trained then to obtain ideal binary masks.

  \textbf{Teacher-student paradigm.}
  Teacher-student learning paradigm \cite{Ashok_2017, Wu_2017} has appeared as a solution to overcome the problem of insufficient labeled training data in MIR.
  Since manual labeling is a time-consuming tasks, the teacher-student paradigm explores the use of unlabeled data for supervised problems.
  The two main agents of this paradigm are: the ‘teacher’ and the ‘student’.
  The \textit{teacher} is trained with labels of well known ground truths datasets (often manually annotated).
  It is then used to automatically label unlabeled data on a (usually) larger dataset.
  These new labels (the one given by the teacher) are the ones used for training the \textit{student(s)}.
  Student(s) indirectly acquire(s) the desired knowledge by mimicking the ``teacher behaviour".
  This model has achived great results for tasks in speech recognition \cite{Watanabe_2017} and multilingual models \cite{Cui_2017}.
  It has also been proved that student(s) can achieve superior performances than the teacher \cite{Wu_2017, Cui_2017}.

  %
  %

  \section{Singing voice dataset: creation}
  \subsection{Karaoke resources}
  \label{sec:karaoke}
  Outside the MIR community there are rich sources of information that can be explored.
  One of these sources is Karaoke video games that fit exactly our requirements.
  In these games, users have to sing along with the music to win points according to their singing accuracy.
  To measure their accuracy, the user melody is compared with a reference timing note (that has fine time and frequency).
  Hence, large datasets of time-aligned note and lyrics exist.

  Such datasets can be found as open-source.
  Nowadays, there are several active and big karaoke open-source communities.
  In those, non-expert users exchange text files containing lyrics and melody annotations.
  However there is no further professional revision.
  Each file contains all the necessary information to describe a song:
  \begin{itemize}
  	\setlength\itemsep{0em}
  	\item the sequence of triplets \{time, musical-notes, text\},
  	\item the {\tt \footnotesize offset\_time} (start of the sequence) and {\tt \footnotesize frame rate} (annotation time-grid),
  	\item the {\tt \footnotesize song\_title} and the {\tt \footnotesize artist\_name}.
  \end{itemize}
  We refer to Table \ref{table:terms} for the definition of all the terms we use.
  These annotations can be transformed to get the time and note frequencies as seen Figure \ref{fig:ex_annotation}.

  We were able to retrieve 13339 karaoke annotation files.
  Although this information is outstanding for our community, it presents several problems that have to be solved:
  \begin{description}[.15cm]
  	\item[Global.] 
  	When performing  the annotation, users can choose the \textit{audio file} they want.
  	The problem is that only the {\tt \footnotesize song\_title} and {\tt \footnotesize artist\_name} are provided.
    This combination might refer to different audio versions (studio, radio edit, live, remix, etc.).
  	Consequently, we do not know which audio version has been used.
  	Annotations made for a version do not work for another.
  	Besides, even if the correct audio is known, annotations may not perfectly fit it.
  	As a result annotations must be adapted.
  	This is done by modifying the provided {\tt \footnotesize offset\_time} and {\tt \footnotesize frame rate}.
  	These issues are not problematic for karaoke-users but critical To the automatic creation of a large dataset for MIR research.
  	\item[Local.]
  	It refers to errors due to fact the that users are non-professionals.
  	It covers local alignment problems of particular lyric blocks, text misspellings or note mistakes.
  \end{description}

  In this paper we only focus on global problems leaving the local ones for future works.\newline

  \begin{figure*}[ht]
  	\centerline{
  		\includegraphics[width=\textwidth]{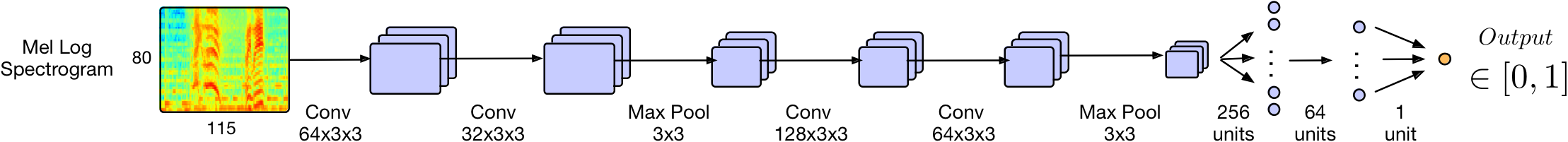}}
  	\caption{Architecture of our Singing Voice Detection system using a ConvNet.}
  	\label{fig:convNet_voice_detection}
  \end{figure*}

  \textbf{\WASABI} is a semantic database of song knowledge gathering metadata collected from various music databases on the Web \cite{meseguerbrocal_2017}.
  In order to benefit from the richness of this database, we first linked each annotation file to Wasabi.
  To that end, we connected a specific {\tt \footnotesize song\_title} and {\tt \footnotesize artist\_name} with all possible corresponding audio versions (studio, radio, edit, live, remix, etc.).
  The \WASABI also provides lyrics in a text only annotations (grouped by lines and paragraphs).
  Using the two lyrics representations (note-based annotations and text only annotations), we created four levels of granularity: notes, words, lines and paragraphs.
  Finally, \WASABI also provides extra multimodal information such as cover images, links to video clips, metadata, biography, expert notes, etc.

  \subsection{Retrieving audio candidates}
  \label{sec:linking_auido}


  Our input is an annotation file connected to the \WASABI database.
  This database provides us with the different existing versions (studio, radio, edit, live, remix, etc.) for a {\tt \footnotesize song\_title} and {\tt \footnotesize artist\_name} combination.
  Knowing the possible versions, we then automatically query YouTube\footnote{We use \url{https://github.com/rg3/youtube-dl}} to get a set of audio candidates.
  We now need to select among the set of audio candidates the one corresponding to the annotation file.

  \subsection{Selecting the right audio from the candidate and adapting annotation to it}
  \label{sec:selecting}

  \MASK{
  	In order to discover its best alignment (hence its best value) for each candidate, annotations are modified by changing its {\tt offset\_time} and {\tt frame rate}.
  	The candidate with the highest value is our final audio track.
  	The critical point here is to define a proper distance measurement.
  	This distance is not only important in recovering the right audio and finding the best alignment but also in determining if annotations are accurate enough to the best audio files we can regain, thereby filtering imprecise annotations.
  }

  Each audio candidate is compared to the reference annotation file.
  We do this by measuring a distance between both and keeping the one with the largest value.


  Audio and annotations live in two different representation spaces that cannot be directly compared.
  In order to find a proper distance, we need to transform them to a common representation space.
  Two directions were studied:

  \begin{description}[.15cm]
  	\item[Annotations as audio.]
  	We have explored lyrics synchronization techniques \cite{Fujihara_2012} but their complexity and phonetic model limitations prevent us to use them.
  	As annotations can be transformed into musical notes, score alignment approaches \cite{Cont_2007, Soulez_2003} seem a natural choice.
  	However, due to missing information in the corresponding score (we only have the score of the vocal melody) these systems failed.
  	We then tried to reduce the audio to the vocal melody (using Melodia \cite{Salamon_2012}) and then align it to the vocal melody score but this also failed.
  	Consequently, we did not persist in this direction.

    %
    %
    %

  	\item[Audio as annotations.]
  	The idea we develop in the remainder is the following.
  	We convert the audio track to a singing-voice probability $\hat{p}(t)$ over time $t$.
  	This sequence has value $\hat{p}(t) \rightarrow 1$ when voice is present at time $t$ and $\hat{p}(t) \rightarrow 0$ otherwise.
  	This probability is computed from the audio signal using a \textbf{Singing Voice Detection} (SVD) system described below.
  	We name this function \textbf{predictions}.
  	Similarly, the sequence of annotated triplets \{time, musical-notes, text\} can be mapped to the same space: $avs(t)=1$ when a vocal note exists at $t$ and $avs(t)=0$ otherwise.
  	We name this function \textbf{annotation voice sequence}.
  \end{description}

  \textbf{Singing Voice Detection system.}
  Our system is based on the deep Convolutionnal Neural Network proposed by \cite{Schluter_2015}.
  The audio signal is first converted to a sequence of patches of 80 Log-Mel bands over 115 time frames.
  Figure \ref{fig:convNet_voice_detection} shows the architecture of the network.
  The output of the system represents the singing voice probability for the center time-frame of the patch.
  The network is trained on binary target using cross-entropy loss-function, ADAMAX optimizer, mini-batch of 128, and 10 epochs.
  \newline

  \textbf{Cross-correlation.}
  To compare audio and annotation, we simply compare the functions $\hat{p}(t)$ and $avs(t)$.
  As explained before, the annotation files also come with a proposed {\tt \footnotesize offset\_time} and {\tt \footnotesize frame rate}.
  We denote them by $O$ and $Fr$ in the following.
  The alignment between $\hat{p}(t)$ and $avs(t)$ depends on the correctness of $O$ and $Fr$ values.
  We will search around $O$ and $Fr$ to find the best possible alignment.
  We denote by $o$ the correction to be applied to $O$ and by $fr$ the best $Fr$.
  Our goals are to:

  \begin{enumerate}
  	\item find the value of $o$ and $fr$ that provides the best alignment between $\hat{p}(t)$ and $avs(t)$,
  	\item based on this best alignment, deciding if $\hat{p}(t)$ and $avs(t)$ actually match each other and establishing if the match is good enough to be kept.
  \end{enumerate}

    \begin{figure*}[ht!]
    	\centerline{
    		\includegraphics[width=\textwidth]{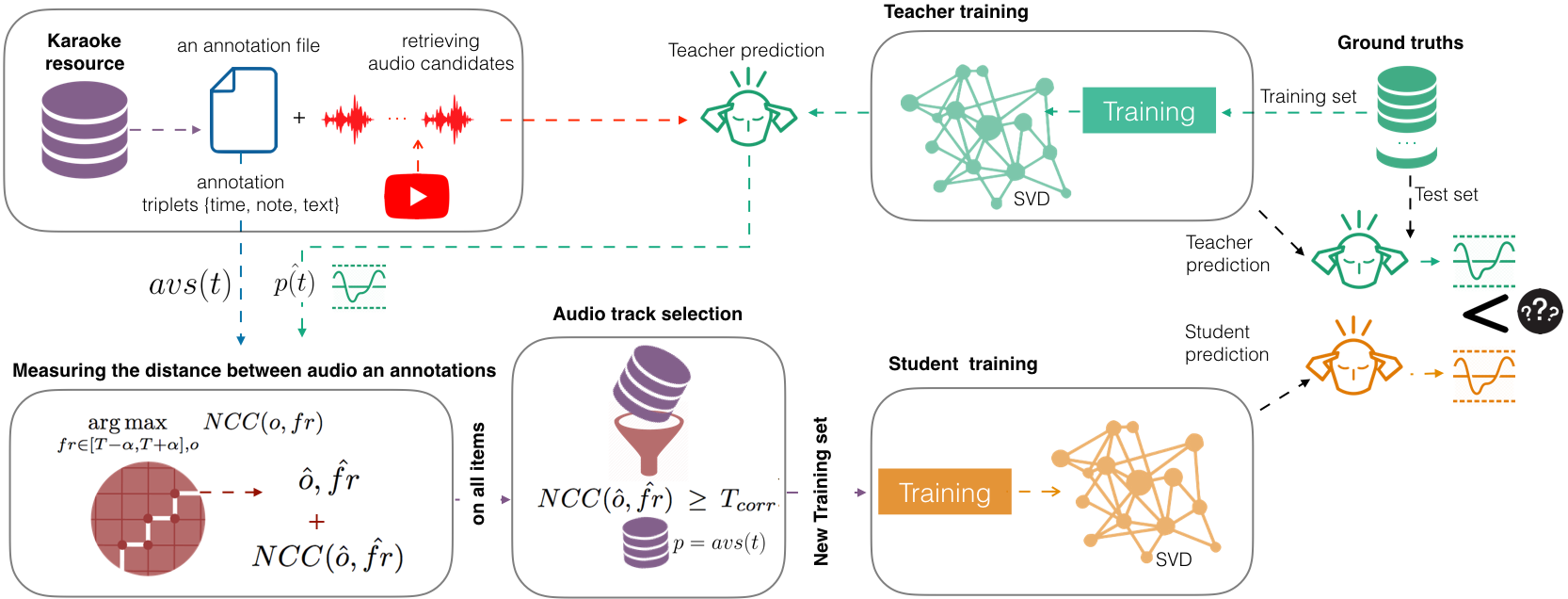}}
    	\caption{Singing Voice Dataset creation using a teacher-student paradigm.}
    	\label{fig:teacher-student}
    \end{figure*}

  Since we are interested in a global matching between $\hat{p}(t) \in [0,1]$ and $avs(t) \in \{0,1\}$ we use the normalized cross-correlation (NCC) as distance\footnote{Matches between $\hat{p}(t)$ and $avs(t)$ can also be found using Dynamic Time Warping (DTW).
  However, we found its application not successfull for our purpose.
  Indeed, DTW computes local warps that does not respect the global structure of the user annotations.
  In addition, its score is not normalized preventing its use for matches selection.}:
  \begin{equation*}
    NCC(o, fr) = \frac{\sum_t avs_{fr}(t-o) \hat{p}(t)}{\sqrt{\sum_t avs_{fr}(t)^2}  \sqrt{\sum_t \hat{p}(t)^2}}
  \end{equation*}

  The NCC provides us directly with the best $\hat{o}$ value.
  This value directly provides the necessary correction to be applied to $O$ to best align both sequences.

  To find the best value of $fr$ we compress or stretch annotation by changing the grid size.
  This warp is constant and respect the annotation structure.
  We denote it as $avs_{fr}(t)$.
  The optimal $fr$ value is computed using a brute force approach, testing the values of $fr$ around the original $Fr$ in an interval controlled by $\alpha$ (we use $\alpha = Fr*0.05$):
  \begin{equation*}
  \hat{fr}, \hat{o} = \argmax_{fr \in [Fr-\alpha, Fr+\alpha] , o } NCC(o, fr)
  \end{equation*}

  Our final score is given by $NCC(\hat{o}, \hat{fr})$.

  The audio is considered as good match the annotation if $NCC(\hat{o}, \hat{fr}) \geq T_{corr}$.
  The value of $T_{corr}$ has been found empirically to be $T_{corr}=0.8$.
  For a specific annotation, if several audio candidate tracks have a value $NCC \geq T_{corr}$, we only keep the one with the largest value.
  $T_{corr}=0.8$ is quite restrictive but even if we may loose good pairs we ensure that those we keep are well aligned.
  When an audio match is found, the annotations are adapted to it using $\hat{fr}$ and $\hat{o}$.\newline


  \textbf{Necessity to improve the Singing Voice Detection system.}
  The score $NCC$ proposed above strongly depends on the quality of $\hat{p}(t)$ (the prediction provided by the \textbf{Singing Voice Detection (SVD)} system).
  Small differences in predictions lead to similar $NCC(\hat{o}, \hat{fr})$ values but very different alignments.
  While the predictions of the baseline SVD system are good enough to select the correct audio candidates (although there are still quite a few false negatives), it is not good enough to correctly estimate $\hat{fr}$ and $\hat{o}$.
  As improving the SVD system the number of false negatives will be reduced and we will also find better alignments.
  We hence need to improve our SVD system.

  The idea we propose below is to re-train the SVD system using the set of candidates audio that match the annotations.
  This is a much larger training set (around 2000) than the one used to train the baseline system (around 100).
  We do this using a teacher-student paradigm.

  \subsection{Teacher-Student}
  \label{sec:teacher-student}

  Our goal is to improve our Singing Voice Detection (SVD) system.
  If it becomes better, it will find better matches and align more precisely audio and annotations.
  Consequently, we will obtain a better DALI dataset.
  This larger dataset can then be used to train a new SVD system which again, can be used to find more and better matches improving and increasing the DALI dataset.
  This can be repeated iteratively.
  After our first iteration and using our best SVD system, we reach 5358 songs in the DALI dataset.

  We formulate this procedure as a Teacher-Student paradigm.
  The processing steps of the whole Singing Voice Dataset creation is summarized in Figure~\ref{fig:teacher-student}.

  \begin{description}[.15cm]
  	\item[Upper left box.]
  	We start from Karaoke resources that provide our set of annotation files.
  	Each annotation file defines a sequence of triplets \{time, note, next\} that we convert to an annotation voice sequence $avs(t)$.
  	For each annotation file, we retrieve a set of audio candidates.

  	\item[Upper right box.]
  	We independently trained a first version of the SVD system (based on ConvNet) using the training set of a ground truth labeled dataset as provided by the Jamendo\cite{Ramona_2008} or MedleyDB\cite{bittner_2014} datasets. We call this first version the \textbf{teacher}.

  	\item[Upper middle part.]
  	This teacher is then applied  on each audio candidate to predict $\hat{p}(t)$.

  	\item[Lower left box.]
  	We measure the distance between $avs(t)$ and $\hat{p}(t)$ using our cross-correlation method.
  	It allows us to find the best audio candidate for an annotation file and the best alignment parameters $\hat{fr}, \hat{o}$.

  	\item[Lower middle box.]
  	We select the audio annotation pairs for which $NCC(\hat{o}, \hat{fr}) \geq T_{corr} =  0.8$.
  	The set of selected audio tracks forms a new training set.

  	\item[Lower right box.]
  	This new set is then used to train a new SVD systems based on the same CNN architecture. This new version is called the \textbf{student}.
  	To this end, we need to define the target $p$ to be minimized in the loss $\mathcal{L}(p,\hat{p})$.
  \end{description}
  There are three choices:
  \begin{itemize}
  	\setlength\itemsep{0em}
  	\item[a)] we use as target $p$ the predicted value $\hat{p}$ given right by the \textbf{teacher} (usual teacher-student paradigm).
  	\item[b)] we use as target $p$ the value $avs$ corresponding to the annotations after aligning them using $\hat{fr}$ and $\hat{o}$.
  	\item[c)] a combination of both, keeping only these frames for which $\hat{p}(t)=avs(t)$.
  \end{itemize}

  Up to now and since the $avs$ have been found more precise than the $\hat{p}$ we only investigated option \textbf{b)}.

  We compare in the following part the results obtained using different teachers and students.

  \subsubsection{Validating the teacher-student pardigm}
  \label{sec:experiments}

  In this part, we demonstrate that the students trained on the new training-set actually perform better than the teacher trained on the ground-truth label dataset.


  \newcommand{\J}[0]{\textit{Jamendo}}
  \newcommand{\M}[0]{\textit{MedleyDB}}
  \newcommand{\both}[0]{\textit{J+M}}

  \begin{description}[.15cm]
  	\item[Ground-truth datasets:]
  	We use two ground-truth labels datasets: \J \cite{Ramona_2008} and \M \cite{bittner_2014}.
  	We created a third dataset by merging \J~  and \M~ named as \both.
  	Each dataset is split into a train and a test part using an artist filter (the same artist cannot appear in both).

  	\item[Teachers:]
  	With each ground-truth datasets we trained a teacher using only the training part.
  	Once trained, each teacher is used to select the audio matches as described in Section \ref{sec:selecting}.
    As a result, we produce three new training sets.
  	They contains 2440, 2673 and 1596 items for the teacher \both, \J~and \M~ respectively.
  	The intersection of the three sets (not presented here) indicates that 89.8 \% of the tracks selected using the \M~teacher are also present within the tracks selected using the \both~teacher or the \J~teacher.
  	Also, 91.4 \% of the tracks selected using the \J~teacher are within the tracks selected using the \both~teacher.
  	It means that the three teachers agree most of the time on selecting the audio candidates.

  	\item[Students:]
    We train three students using the audio and the $avs$ value of the new training sets.
    Even if there is a large audio files overlap within the training sets, their alignment (and therefore the $avs$ value) is different.
    The reason to this is that each teacher gets a different $\hat{p}$ which results in different $\hat{fr}, \hat{o}$ values.


  \end{description}

  \subsubsection{Results}

  We evaluate the performances of the various teachers and students SVD systems using the test parts of \J~(J\_test) and \M~(M\_test).
  We measure the quality of each SVD system using the frame accuracy i.e. average value over all the tracks of the test set.

  Results are indicated in Table~\ref{table:teacher-student}.
  In this table, e.g. ``Student (Teacher\_J\_train) (2673)" refers to the student trained on the 2673 audio candidates and the $avs$ values computed with the Teacher trained on \J~train set.

  %

  \begin{table}[ht]
  	\centering
  	\caption{Performances of the teachers and students using the various datasets. Number of tracks in brackets.}
  	\label{table:teacher-student}
  	\small
  	\begin{tabular}{| c | c  | c |}
  		\hline
  		\backslashbox{SVD system}{Test\_set} 			& J\_test (16) 		& M\_test (36)   \\
  		\hline
  		\hline
  		Teacher\_J\_train (61)         					& 87\%     			& 82\% \\
  		Student (Teacher\_J\_train)  (2673)    			& 82\%       		& 82\% \\
  		\hline
  		Teacher\_M\_train (98)        					& 76\%      		& 85\% \\
  		Student (Teacher\_M\_train) (1596)    			& 80\%     			& 84\%  \\
  		\hline
  		Teacher\_J+M\_train (159)   						& 82\%      		& 82\%  \\
  		Student (teacher\_J+M\_train)     (2440) 			& 86\%   			& 87\%  \\
  		\hline
  	\end{tabular}
  \end{table}

  \newcommand{\JT}[0]{\textit{Teacher\_J\_train~}}
  \newcommand{\MT}[0]{\textit{Teacher\_M\_train~}}
  \newcommand{\JMT}[0]{\textit{Teacher\_J+M\_train}}

\begin{description}[.15cm]\setlength{\parindent}{1pc}
  \setlength{\parskip}{.005cm plus0mm minus0mm}
      \item[Performance of the teachers.]
      We first test the teachers.
      \JT obtains the best results on J\_test (87\%).
      \MT obtains the best results on M\_test (85\%).
      In both cases, since training and testing are performed on two parts of the same dataset, they share similar audio characteristics.
      These results are artificially high.
      To best demonstrate the generalization of the trained SVD systems, we need to test them in a cross-dataset scenario, namely train and test in different datasets.

      Indeed, in this scenario the results are quite different.
      Applying \JT on M\_test the results decreases down to 82\% (a 5\% drop).
      Similarly when applying \MT on J\_test the results decreases down to 76\% (a 9\% drop).
      Consequently, we can say that the teachers do not generalize very well.

      Lastly, the \JMT~ trained on J+M\_train actually performs worse on both J\_test (82\%) and M\_test (82\%) than their non-joined teacher (87\% and 85\%).
      These results are surprising and remain unexplained.

      \item[Performance of the students.]
      We now test the students.
      It is important to note that students are always evaluated in a cross-dataset scenario since the DALI dataset (on which they have been trained) does not contain any track from \J~or \M.
      Hence, there is no possible overfitting for those.
      Our hypothesis is that students achieve better the results than the teachers because they have seen more data.
      Especially, we assume that their generalization to unseen data will be better.

      This is true for the performances obtained with the student based on \MT.
      When applied to J\_test, it reaches 80\% which is higher than the performances of the \MT directly (76\%).

      This is also true for the performances computed with the student based on \JMT.
      When applied either to J\_test or M\_test, it reaches 86.5\% (86\% on \J~ and 86\% on \M) which is above the \JMT~ (82\%).
      Also, 86.5\% is similar or above the results obtained with \JT on J\_test (87\%) and \MT on M\_test (85\%).
      This is a very interesting result that demonstrates the generalization of the student system whichever data-set it is applied to.
      The \textbf{student} based on \textbf{\JMT} is the one used for defining the final 5358 songs of the DALI dataset.

      However, the performances obtained with the student based on \JT applied to M\_test (82\%) do not improve over the direct use of the \JT (82\%).

      \item[On alignment.]
      Not explained in this paper is the fact that the $\hat{fr}$ and $\hat{o}$ values computed with the students are much better (almost perfect) than the ones obtained with the teacher.
      However, we cannot measure it precisely since DALI dataset does not have ground-truth label annotations to that end.
      Indeed, the goal of this paper is exactly to obtain such annotations automatically.
  \end{description}

  \section{Singing voice Dataset: access}
  \label{sec:dataset_description}
  The DALI dataset can be downloaded at {\footnotesize \url{https://github.com/gabolsgabs/DALI}}.
  There, we provide the detailed desciption of the dataset as well as all the necessary information for using it.
  DALI is presented under the recommendation made by \cite{Peeters_2012} for the description of MIR corpora.
  The current version of DALI is 1.0. Future updates will be detailed in the website.


  \section{Conclusion and future works}
  \label{sec:discussion}

  In this paper we introduced DALI, a large and rich multimodal dataset containing 5358 audio tracks with their time-aligned vocal melody notes and lyrics at four levels of granularity.

  We explained our methodology where dataset creation and learning models interact using a teacher-student paradigm benefiting one-another.
  From manual karaoke user annotations of time-aligned lyrics and notes, we found a set of matching audio candidates from the Web.
  To select and align the best candidate, we compare the annotated vocal sequence (corresponding to the lyrics) to the singing voice probability (obtained with a ConvNet).
  To improve the latter (and therefore obtain a better selection and alignment) we applied a teacher-student paradigm.

  Through an experiment, we proved that the students outperform the teachers notably in a cross-dataset scenario, when train-set and test-set are from different datasets.

  It is important to note that the results of the students are higher than the teacher ones, even if they have been training on imperfect data.
  In our case, we showed that, in the context of deep learning, it is better to have imperfect but large dataset rather than small and perfect ones.
  However, other works went in the opposite direction~\cite{Amatriain2017}.

  \textbf{Future work.}
  We have only performed the teacher-student iteration once.
  In next works will use the results of the first student generations to train a second student generations.
  This will define a new DALI dataset.
  We plan to quantitative measure the quality of $\hat{o}, \hat{fr}$ and to continue exploring the alignments between note annotations and the audio.
  Currently, we trained our student using as target $p=avs$, which do not transfer directly the knowledge of the teacher.
  We will explore other possibilities of knowledge transfer using other targets (points a) and c) in Section \ref{sec:teacher-student}) as well as the local problems describe at Section \ref{sec:karaoke}.

\textbf{Acknowledgement.}
This research has received funding from the French National Research Agency under the contract ANR-16-CE23-0017-01 (WASABI project).

  \newpage
  \bibliography{ismir2018.bib}
\end{document}